\documentclass[conference]{IEEEtran}
\usepackage[numbers]{natbib}
\usepackage{graphicx}
\usepackage{bm}
\usepackage{etoolbox}
\makeatletter
\patchcmd{\@makecaption}
 {\\}
 {.\ }
 {}
 {}

\ifCLASSINFOpdf
\else
\fi
\usepackage{amsmath}
\usepackage{amssymb}

\DeclareMathOperator{\sig}{sig}

\ifCLASSOPTIONcompsoc
\usepackage[caption=false,font=normalsize,labelfont=sf,textfont=sf]{subfig}
\else
\usepackage[caption=false,font=footnotesize]{subfig}
\fi
\begin{document}

\title{Super-Resolution for Practical Automated\\
Plant Disease Diagnosis System}

\author{\IEEEauthorblockN{Quan Huu Cap\IEEEauthorrefmark{1}, Hiroki Tani\IEEEauthorrefmark{1}, Hiroyuki Uga\IEEEauthorrefmark{2}, Satoshi Kagiwada\IEEEauthorrefmark{3} and Hitoshi Iyatomi\IEEEauthorrefmark{1}}
\IEEEauthorblockA{Email: \{huu.quan.cap.78, hiroki.tani.2j\}@stu.hosei.ac.jp uga.hiroyuki@pref.saitama.lg.jp \{kagiwada, iyatomi\}@hosei.ac.jp}
\IEEEauthorblockA{\IEEEauthorrefmark{1}Applied Informatics, Graduate School of Science and Engineering, Hosei University, Tokyo, Japan}
\IEEEauthorblockA{\IEEEauthorrefmark{2}Saitama Agricultural Technology Research Center, Saitama, Japan}
\IEEEauthorblockA{\IEEEauthorrefmark{3}Clinical Plant Science, Faculty of Bioscience and Applied Chemistry, Hosei University, Tokyo, Japan}}

\maketitle

\begin{abstract}
Automated plant diagnosis using images taken from a distance is often insufficient in resolution and degrades diagnostic accuracy since the important external characteristics of symptoms are lost. In this paper, we first propose an effective pre-processing method for improving the performance of automated plant disease diagnosis systems using super-resolution techniques. We investigate the efficiency of two different super-resolution methods by comparing the disease diagnostic performance on the practical original high-resolution, low-resolution, and super-resolved cucumber images. Our method generates super-resolved images that look very close to natural images with 4$\times$ upscaling factors and is capable of recovering the lost detailed symptoms, largely boosting the diagnostic performance. Our model improves the disease classification accuracy by 26.9\% over the bicubic interpolation method of 65.6\% and shows a small gap (3\% lower) between the original result of 95.5\%.\\
\end{abstract}

\begin{IEEEkeywords}
super-resolution, deep learning, automated plant disease diagnosis, cucumber plant diseases
\end{IEEEkeywords}


\section{Introduction}
Diagnosing plant diseases is generally conducted by visual examination through experts. Thus, it is often time-consuming and expensive tasks. Increasing number of computer-based diagnostic methods have been proposed to effectively prevent plant diseases and reduce the loss of crop yield. These researches dramatically increased with the innovation of deep learning techniques. Early studies using these techniques were to analyze one leaf image and yield the diagnosis result [1--5]. Recently, some of more sophisticated methodologies detect multiple targets of plant to be measured [6] or diagnosed [7, 8] simultaneously from relatively wide-shot images (i.e. distance from camera to target is up to approximately 2m). In the former, they counted and measured plant stalk for supporting robot harvesting. In the latter, they investigated on-site tomato leaf images [7] and wheat images [8] taken from slightly wide-shot, respectively. Despite those systems achieving excellent diagnostic performance, they still suffer from a limitation that their target images are considerably more narrow range than fixed-point observation camera images which is expected to be in practical applications. We have experienced that the accuracy is insufficient when solely applying these methods to wide-angle images. The wide-angle images on the practical agricultural site contain numerous of objects to be detected, many which are visually similar and overlapping with one another. Thus, the system needs to make a diagnosis with these slight differences. In addition, key techniques used in the abovementioned systems were originally designed for general object recognition; they implicitly assume that the physical appearances of the recognition target is notably different among categories. However, this assumption does not apply to practical plant diagnosis task on wide-angle images. We believe one of the main reasons for this issue is the lack of resolution on the to-be-diagnosed targets in wide-angle images. The images taken from a distance often include smaller and lower resolution objects. Thus, it would decrease the performance of diagnosis systems since additional localization must be performed beforehand. The same issue is reported in a study of the end-to-end disease diagnosis system for wide-angle cucumber images [9]. They admitted that the small leaf size and low-quality input images (low-resolution, blur, poor camera focus, etc.) could significantly reduce their disease diagnostic performance. We should note that this problem can be avoided by using a high-resolution camera device to capture the image, but it is generally expensive to deploy in practice.\\
\indent We believe the solution for this problem is recovering the high-frequency component of images by applying the super-resolution (SR) methods. The SR methods had required multiple images to attain a certain level of performance [10]. But thanks to the modeling power of convolutional neural networks (CNN), recently SR methods based on only single image, so-called single image super-resolution (SISR) have been proposed and shown excellent performance [11, 12, 13]. The pioneer work on SR was the super-resolution convolutional neural network (SRCNN) [11]. The SRCNN directly learned the mapping function between low and high-resolution images, providing an end-to-end training manner. They achieved significant improvement over conventional SR methods. The SRGAN [12] was then proposed as the first SR method using the advantage of the generative adversarial networks (GAN) [14]. Furthermore, the enhanced SRGAN (ESRGAN) [13] including the residual-in-residual dense block (RRDB) and the applied relativistic average GAN (RaGAN) [15] as the key components was proposed and outperformed SRGAN in term of perceptual quality.\\
\begin{figure*}[t]
 \centering
 \includegraphics[width=\linewidth]{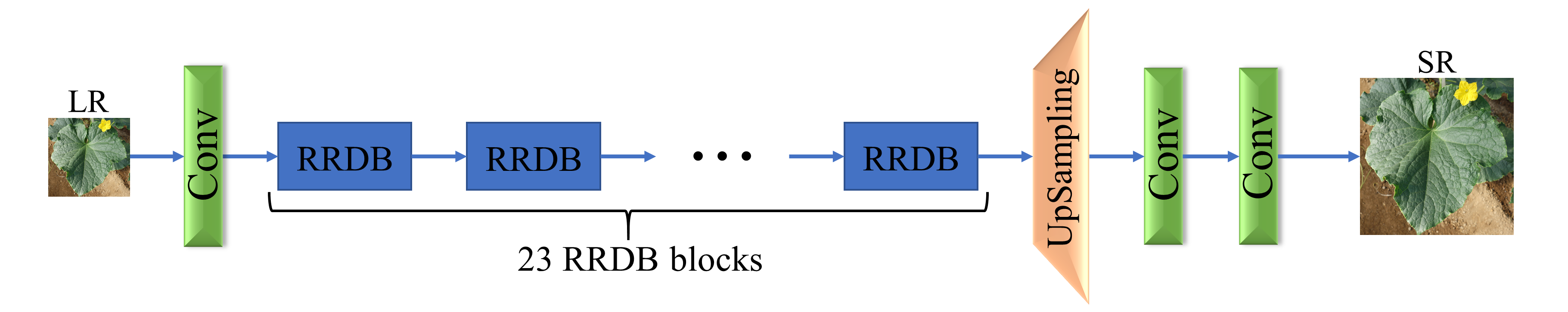}
  \caption{The generator $G$ consists of 23 RRDB blocks followed by the up-sampling and convolutional layers to generate the SR images.}
  \label{fig:the_generator_architecture}
\end{figure*}
\indent Although the SR techniques have been widely used in many fields, few applications to the agricultural sector have been seen so far [16, 17]. In [16], they proposed an adaptive based image SR method to enhance resolution of disease leaf images. They claimed this could provide pathologists a better visual assessment on the infected leaves, but so far, they have not evaluated the effects of their SR method for automated plant diagnosis. The use of SRCNN in the automated plant diagnosis has firstly been noted in [17]. Authors compared the disease classification performance on tomato between the interpolated low-resolution, the estimated SR, and the original high-resolution tomato leaf images. They demonstrated SR using SRCNN boosted the disease detection accuracy by 20\%, achieving 90\%. Although they showed a promising result, we cannot conclude from their result that this diagnostic performance could be used in practice, or SR actually contributed to improving diagnostic accuracy in the practical situation. This is because the result was based on an in-lab environment since they used the PlantVillage [18] dataset. Each leaf image in the dataset is taken in an ideal situation; they are manually cropped and placed on a uniform background. In practice, on-site leaf images appear more complex with different symptoms, varying in backgrounds and are affected with various photographic conditions. It is already known that the diagnosis system trained with these images showed extremely high diagnostic accuracy, but the performance was devastatingly low when applied to real on-site images [2].\\
\indent To the best of our knowledge, there is no literature to apply the SR methods on plant disease diagnosis in practical situations. In this paper, we propose an effective pre-processing method for improving the performance of automated plant disease diagnosis systems from on-site images using super-resolution techniques.

\section{Method}
We introduce two SR models, $M_{pix}$ and $M_{feat}$, and investigate how they improve the practical plant diagnostic accuracy. The $M_{pix}$ is similar to SRCNN and its loss function $L_{pix}$ is to minimize the differences between generated SR and HR images in the pixel space. On the other hand, the $M_{feat}$ is based on ESRGAN and its loss function $L_{feat}$ is to minimize the differences between those images in the feature space instead of pixel space [19]. According to our preliminary experiments, the $M_{pix}$ could recover the important disease features and reduce unexpected noises since it produces smoothed SR images. While the $M_{feat}$ could recover more high-frequency details and generate the perceptually more pleasing results. However, in very few cases, $M_{feat}$ yielded some tiny artifacts on the generated leaf surface. We expect both methods could improve the diagnosis performance on plant disease images and would like to determine which method is better suited for our purpose.\\
\indent For evaluation, we use the multiple diseases cucumber dataset in [20] because of its practicality. Their images were taken on site and consists of a wide variety of backgrounds and lighting conditions. In addition, they were planted in a strictly controlled environment and therefore, all of these images have reliable disease label for gold standard. We develop a disease classifier with this dataset and compare their diagnostic performance among different resolution images, i.e. the down-sampled and interpolated low-resolution (LR) images, generated SR images, and the original high-resolution (HR) leaf images. All the interpolated and SR images will be generated with 4$\times$ upscaling factor from LR images (i.e. the ratio size of those images is 1:16).
\begin{figure*}[t]
 \centering
 \includegraphics[width=\linewidth]{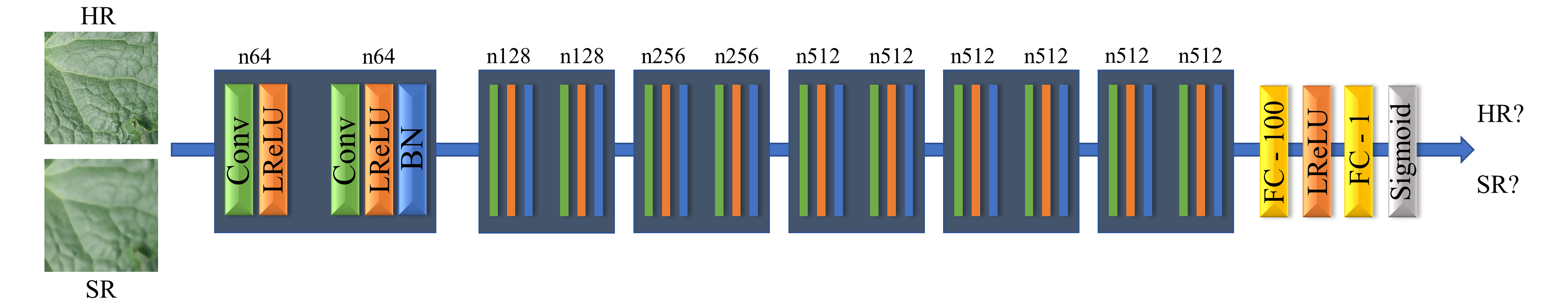}
  \caption{The discriminator $D$ consists of six $conv\_block$ with corresponding number of feature maps $n$.}
  \label{fig:the_discriminator_architecture}
\end{figure*}
\subsection{Network Architectures}
The $M_{pix}$ is composed of one generator CNN model for generating SR images. On the other hand, the $M_{feat}$ is composed of two CNN models; the generator $G$ and the discriminator $D$ as referred in the ESRGAN. Note here that the generator $G$ generates SR images and the discriminator $D$ distinguishes generated SR images from the HR images. We call the generator CNN model of $M_{pix}$ as $G_{pix}$ and the model of $M_{feat}$ as $G_{feat}$, respectively. Both $G_{pix}$ and $G_{feat}$ have the same network architecture as for the generator $G$ of ESRGAN reported in the original literature [13]. In the $M_{feat}$, two networks $G_{feat}$ and $D$ are trained together to solve the adversarial min-max problem.
\subsubsection{The Generator}The generator $G$ is composed of multiple residual-in-residual dense blocks (RRDB) [13]. An RRDB consists of three dense blocks [21]; each dense block has five densely connected convolutional layers. The three dense blocks are connected to the main path of the network in a residual manner [22]. In this work, we use the generator $G$ consisting of 23 RRDB blocks, resulting in a total of 115 convolutional layers. Fig. 1 illustrates the generator $G$ architecture used in our experiments.
\subsubsection{The Discriminator}Our discriminator $D$ is designed under the architecture guidelines for stable deep convolutional GANs in [23]. The difference from the discriminator in ESRGAN is its input size. We design our discriminator deeper to accept larger input size of 192$\times$192 compared to the original 128$\times$128. We found that the larger input size could help the network learn useful information. The architecture of our discriminator $D$ is illustrated in Fig. 2. Our discriminator consists of six convolutional blocks ($conv\_block$) followed by two fully-connected (FC) layers and has no max-pooling layers. We define our $conv\_block$ as a block of two convolutional layers. Each convolutional layer has its kernel size $k$, number of feature maps $n$ and stride $s$. At each $conv\_block$, we use $k_1=3$, $s_1=1$ for the first and $k_2=4$, $s_2=2$ for the second convolutional layer. The number of $n$ is different on each $conv\_block$. The first fully-connected layer has 100 units while the last layer contains a single unit and a sigmoid activation function. We use the leaky rectified linear function (LReLU) [24] with $\alpha=0.2$ as the activation function for all layers except for the last layer. Batch normalization (BN) [25] is applied from the second to the last convolutional layer.
\subsubsection{Loss Functions}The objective of training $M_{pix}$ is to minimize the pixel-wise differences between generated SR images $I_{SR}$ and HR images $I_{HR}$. We choose the loss function of $G_{pix}$ to be minimized as:
\begin{equation}
L_{pix}=\left|I_{HR} - I_{SR}\right|_{1}.
\end{equation}
\indent For the training of $M_{feat}$, the adversarial training between $G_{feat}$ and the discriminator $D$ is applied. The output of our discriminator $D$ is expressed as:
\begin{equation}
D(I_{HR},I_{SR})=\sig(C(I_{HR})-\mathbb{E}_{I_{SR}}\left[C(I_{SR}) \right])
\end{equation}
,where $C(I)$ is the output from the $FC-1$ layer (before the sigmoid layer) of the discriminator $D$ (see Fig. 2); $\mathbb{E}_I[\cdot]$ is the average value of all images in a mini-batch $I$. Note that the $I_{HR}$ and $I_{SR}$ in Eq. (2) can be substituted for each other.
Here, we use the same loss functions for $G_{feat}$ and $D$ as used in ESRGAN. The adversarial loss for the discriminator $L_D$ is defined as:
\begin{multline}
L_D=-\mathbb{E}_{I_{HR}}\left[\log(D(I_{HR},I_{SR}))\right] \\-\mathbb{E}_{I_{SR}}\left[\log(1-D(I_{SR},I_{HR})) \right].
\end{multline}
Therefore, the adversarial loss for generator $L_{G_{feat}}$ is in a symmetrical form:
\begin{multline}
L_{G_{feat}}=-\mathbb{E}_{I_{HR}}\left[\log(1-D(I_{HR},I_{SR}))\right]\\
-\mathbb{E}_{I_{SR}}\left[\log(D(I_{SR},I_{HR})) \right].
\end{multline}
Finally, we represent the total loss $L_{feat}$ for the generator $G_{feat}$ as:
\begin{equation}
L_{feat}=L_{percep}+\lambda L_{G_{feat}}+\eta\left|I_{HR}-I_{SR} \right|_1
\end{equation}
,where $\lambda$, $\eta$ are the coefficients to balance different loss terms. $L_{percep}$ is the perceptual loss in the features space of HR and generated SR images represented in the VGG19 [26] model pretrained with ImageNet dataset [27]. The $L_{percep}$ is defined as:
\begin{equation}
L_{percep}=\left|VGG19_{5\_4}(I_{HR})-VGG19_{5\_4}(I_{SR}) \right|_1.
\end{equation}
Here, $VGG19_{5\_4}$ is the convolution layer before the last max-pooling layer of the VGG-19 model.
\subsection{The Cucumber Diseases Dataset}
All the cucumber leaf images were taken from Saitama Agricultural Technology Research Center, Japan. Each original HR image contains a single cucumber leaf roughly in the center surrounded with various backgrounds. There are seven types of viral (CCYV, CMV, KGMMV, MYSV, PRSV, ZYMV and WMV) and four fungal (Brown spot, Downy mildew, Gray mold, and Powdery mildew) diseases, resulting 11 disease types. For multiple diseases, each case generally has two or three diseases and a total of 13 combinations of multiple infections were labeled from above 11 diseases. The dataset has a total of 48,311 cucumber leaf images consisting of 38,821 single, 1,814 multiple infections, and 7,676 healthy leaves. Given this dataset, we divide the training and testing set into two separate sets. Specifically, the training set has 36,233 images (roughly 75\% of dataset) and the testing set contains 12,078 images (roughly 25\% of dataset). We treat this task as 25-class classification task as well as in [20].
\begin{figure*}[t]
\centering
\subfloat[Bicubic]{\includegraphics[width=0.24\linewidth]{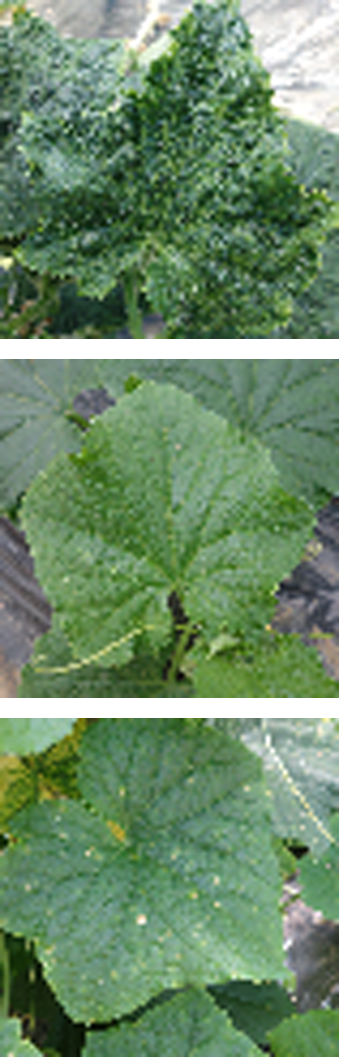}
\label{fig_first_case}}
\hfil
\subfloat[$G_{pix}$]{\includegraphics[width=0.24\linewidth]{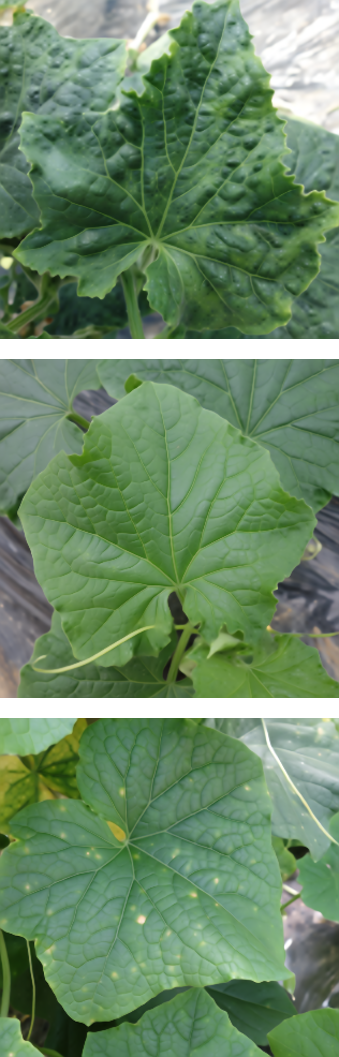}
\label{fig_second_case}}
\hfil
\subfloat[$G_{feat}$]{\includegraphics[width=0.24\linewidth]{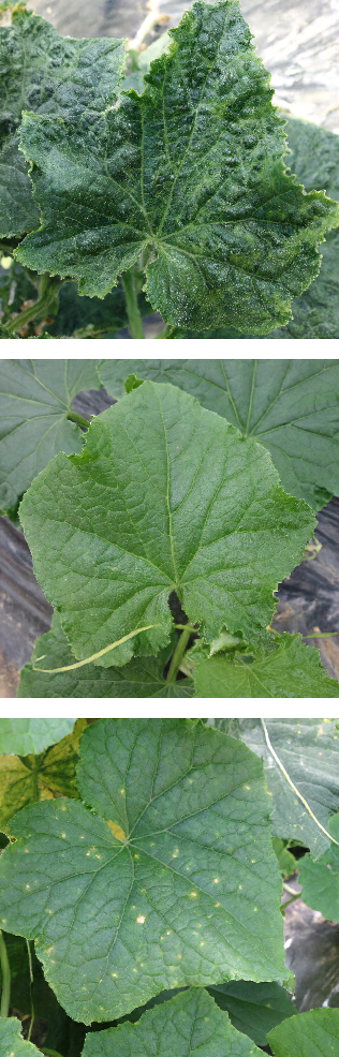}
\label{fig_third_case}}
\hfil
\subfloat[Original]{\includegraphics[width=0.24\linewidth]{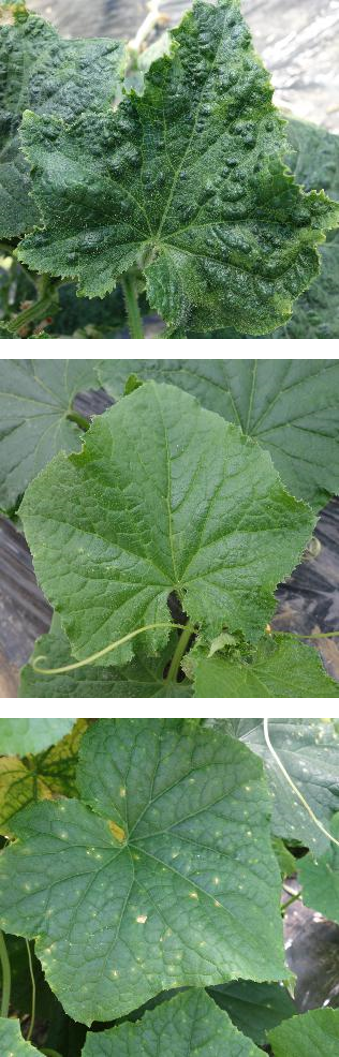}
\label{fig_fourth_case}}
\caption{The visual comparison between the generated and the original images. Super-resolved images from $G_{feat}$ are almost indistinguishable from original ones and have more details compared to other SR methods (bicubic and $G_{pix}$).}
\label{fig_visual_result}
\end{figure*}
\subsection{The CNN Architecture for Diseases Diagnosis}
In order to compare the diagnostic performance between the generated SR images and the original HR images, we train our $CNN_{Diag}$ for diagnosing diseases on the training set (containing 36,233 images). The $CNN_{Diag}$ is the same model as the one proposed in [20]. It accepts the input size of 224$\times$224 and consists of eight convolutional layers followed by two fully-connected layers. We apply ReLU as the activation function for all layers except for the last fully-connected layer and the BN for all convolutional layers. Two fully connected layers have 2,048 nodes each with a dropout [28] ratio of $0.5$. To classify multiple infections, a sigmoid function with tunable thresholds on each node is used in the last layer of the network. The thresholds for each output value were set beforehand by applying grid-search on F1-score result domain in order to deal with the imbalance data problem.
\section{Experiments and results}
\subsection{Training the $CNN_{Diag}$}
We apply the same data augmentation method as used in [20], resulting 36 times larger than the original training dataset. We train our $CNN_{Diag}$ model using Adam optimization [29] with mini-batch size of 128. After 1,000 epochs, we achieved the accuracy on classifying cucumber diseases of 95.5\%.
\subsection{Training the SR models}
We train our $M_{pix}$ and $M_{feat}$ using the same training images used to train the $CNN_{Diag}$. During the training, the HR images are obtained by randomly cropping the training images with a pre-defined size. The LR images are then created by down-sampling from HR images using bicubic interpolation. Both LR and HR images are augmented with random horizontally flip and random 90 degrees rotation on-the-fly.\\
\indent First, we train the $G_{pix}$ with the pixel-wise loss $L_{pix}$ in Eq. (1). The HR images are randomly cropped with the size of 96$\times$96 from training images. Based on our preliminary experiments, the $G_{pix}$ which is trained with smaller HR crop size (96$\times$96) produces better visual results. Mini-batch size is set to 64 and the training is finished after 1,000,000 iterations (roughly 1,780 epochs).\\
\indent Second, we train the $G_{feat}$ with the initial weights from the pre-trained $G_{pix}$. Our $G_{feat}$ is trained along with the discriminator $D$ using the loss functions in Eq. (3) and Eq. (5) with $\lambda=5\times10^{-3}$ and $\eta=10^{-2}$. We set the mini-batch size of 32 images. HR images with size of 192$\times$192 are randomly cropped from training images. Different from the crop size of $G_{pix}$, the bigger HR size (192$\times$192) used in $G_{feat}$ could help to capture more semantic information. We finish adversarial training for $G_{feat}$ after 400 epochs. Note that we use the Adam as the optimizer for training both $G_{pix}$ and $G_{feat}$.
\subsection{Results}
To evaluate the results of our SR models, all 12,078 images from the testing set were resized to the size of 56$\times$56 by bicubic interpolation beforehand.\\
\indent We enlarge LR images with 4$\times$ upscaling factors by using the bicubic interpolation and the pre-trained models $G_{pix}$, $G_{feat}$. The visual comparison between the generated and the original images is shown in Fig. 3. Since LR images are tiny in size (56$\times$56), the generated images by bicubic interpolation are low in quality, blurred and unable to recover the details. The results from $G_{pix}$ are over-smoothed but have much better quality than bicubic interpolated images. On the other hand, the $G_{feat}$ produces more natural images with recovered high-frequency details. They are almost indistinguishable from original images.\\
\indent Table I shows the accuracy of diseases diagnosis using the generated images by bicubic, $G_{pix}$, $G_{feat}$, and original HR images. These results indicate that the $G_{feat}$ with perceptual loss performs the best among the other SR methods with 92.5\% of average diagnosis accuracy on its generated images. Additionally, it shows a small gap between the original result (only 3\% lower), outperforming the low results from bicubic and $G_{pix}$ (65.6\% and 71.8\% respectively).
\begin{table}[t]
\centering
\caption{The quantitative results of disease diagnosis produced by bicubic, $G_{pix}$, $G_{feat}$, and original images}
\label{tab:example}
\begin{tabular}{|c|c|c|c|c|}
\hline
& \textbf{Bicubic} & $\bm{G_{pix}}$ & $\bm{G_{feat}}$ & \textbf{Original}\\
\hline
Accuracy (\%) & 65.6 & 71.8 & \textbf{92.5} & 95.5\\
\hline
\end{tabular}
\end{table}
\section{Discussion}
We investigated the effectiveness of SR methods for improving the performance of automated plant disease diagnosis system on a practical cucumber image dataset. From the results in Fig. 3 and Table I, there is no surprise that the low-quality images generated by bicubic interpolation yield the lowest diagnosis performance (only 65.6\%). As for the result of $G_{pix}$, although they generate a finer visual quality than bicubic interpolation, the diagnosis result did not increase significantly (only 6.2\% higher). Since $G_{pix}$ was trained with the pixel-wise loss, it is not able to recover the high-frequency image components. In this case, recovering the disease symptoms which appear in a high-frequency detailed form is crucial for improving diagnosis systems. Thus, the SR methods that use pixel-wise loss are not suitable for practical on-site disease diagnosis.\\
\indent On the other hand, our $G_{feat}$ shows an outstanding diagnosis result with 92.5\% mean accuracy which dramatically improved 20.7\% from $G_{pix}$ and close to the original high-resolution result. This reinforces our inference that the practical disease symptoms usually appear in the high-frequency detailed form. Recovering the lost detailed disease symptoms could improve the diagnosis performance. Moreover, our preliminary experiment showed that within the first 100 training epochs, our $G_{feat}$ model achieved nearly 89\% of diagnostic performance. This indicates that the perceptual SR method is an effective tool for practical disease diagnosis on LR images.\\
\indent In our experiments, our $G_{feat}$ yielded some tiny artifacts on generated leaf surface for very few cases. Although the effect of these minor noises on the classification accuracy was limited, we are currently investigating to overcome this problem.\\
\indent Although our model achieves an excellent result, we believe it is also somewhat dominated by the $CNN_{Diag}$ model. For improving the practicality of automated disease diagnosis system, we will continue to develop our SR method to work with different magnification scales and apply it into the wide-angle images such as images taken by surveillance cameras in future studies.

\section{Conclusion}
To the best of our knowledge, this paper is the first to propose an effective pre-processing method for automated plant disease diagnosis systems using SR methods. We have achieved a promising diagnosis result under low-quality image conditions on the practical multiple diseases cucumber dataset. From these results, we have confirmed that our SR method with perceptual loss is efficient and suitable for improving the practical disease diagnosis performance.

\section*{Acknowledgment}
This research was partially supported by the Ministry of Education, Culture, Science and Technology of Japan (Grant in Aid for Fundamental research program (C), 17K8033, 2017-2020).

\nocite{*}
\footnotesize{
\bibliographystyle{IEEEtran}
\bibliography{reference}
}
\end{document}